\documentclass[twocolumn,showpacs,amsmath,amssymb,10pt]{revtex4}
\usepackage[T1]{fontenc}
\usepackage[latin9]{inputenc}
\usepackage{color}
\usepackage{amsmath}
\usepackage{amssymb}
\usepackage{graphicx}
\makeatletter
\usepackage{amsfonts}

\newcommand{\ot}{{\,\otimes\,}}
\newcommand{\eA}{e^{(A)}}
\newcommand{\eB}{e^{(B)}}
\newcommand{\rA}{\rho^{(A)}}
\newcommand{\rB}{\rho^{(B)}}
\newcommand{{\Cd}}{{\mathbb{C}^d}}
\newcommand{{\Cn}}{{\mathbb{C}^n}}
\newcommand{{\cp}}{{completely positive\ }}
\newcommand{{\ew}}{{entanglement witness\ }}
\newcommand{{\ews}}{{entanglement witnesses\ }}
\newcommand{{\spa}}{{structural physical approximation\ }}

\def\<{\langle}\def\>{\rangle}

\newtheorem{tw}{Theorem}
\newtheorem{ex}{Example}
\newtheorem{prop}{Proposition}
\newtheorem{cor}{Corollary}
\newtheorem{Rm}{Remark}

\begin{document}
\title{\textbf{Circulant states with vanishing quantum discord}}

\author{Bogna Bylicka and Dariusz Chru\'{s}ci\'{n}ski\\
 Institute of Physics, Nicolaus Copernicus University,\\
 Grudzi\c{a}dzka 5/7, 87--100 Toru\'{n}, Poland}


\begin{abstract}
We analyze a class of 2-qudit circulant states. They define
 generalization of well known  2-qubit X-states. We formulate necessary
 and sufficient criteria for vanishing quantum discord. We illustrate
our analysis by an important subclass of circulant states -- so called
Bell diagonal states.
\end{abstract}


\maketitle

\section{Introduction}

Quantum state of composite system contains both classical and
quantum correlation. Usually by quantum correlations one means
quantum entanglement which provides an essential resource
for quantum information processing such as quantum cryptography,
dense coding, quantum computing \cite{N&Ch,HHHH}.

However, a quantum state of a
composed system may contain other types of nonclassical
correlation even if it is separable (not entangled). For a recent
``catalogue" of nonclassical correlations, see \cite{RE_QD}.  The most
popular measure of such correlations -- quantum discord --
was introduced by Ollivier and Zurek \cite{QD_Zurek} and independently by
Henderson and Vedral \cite{QD_Vedral}.

Recently quantum discord has received increasing attention. It was analyzed in the context of
 broadcasting of quantum states \cite{B1,B2}. Interestingly, it turned out that quantum
discord might be responsible for the quantum computational
efficiency of some quantum computation tasks \cite{QC1, QC2, QC3, QC4}. Moreover, the
 dynamics of discord, both Markovian and non-Markovian,
\cite{M1,M_Vedral, M2, NM1,NM2,NM3,NM4, QD_Dyn1, QD_Dyn2, QD_Dyn3} was analyzed.
Quantum discord was generalized for continuous
variables to study correlations in Gaussian states
\cite{GaussQD_Adesso, GaussQD1}. A geometric measure for quantum discord was introduced in \cite{QD_Zero+Geo} and analyzed in \cite{QD_GeoL}. Finally, an
operational interpretation of quantum discord was provided in
\cite{InterQD_Cavalcanti, InterQD_Datta} -- quantum discord received clear information-theoretic operational
meaning in terms of entanglement consumption in an extended quantum state merging protocol.

Remarkable progress in characterization of set of zero-discord
states was done. Interestingly, it was shown \cite{ZeroQD_Acin} to
have vanishing volume in the set of all states. Actually, this
result holds true for any Hilbert space dimension. It shows that a
generic state of composed quantum system does contain non-classical
correlation. Necessary and sufficient conditions were provided
\cite{QD_Zero+Geo, ZeroQD_Modi,ZeroQD_Datta, ZeroQD_2N}, to
determine states with vanishing discord. Moreover nonlinear
witnesses of discord were introduced \cite{QDWitness1, QDWitness2, QDWitness4}.
For very recent papers analyzing various aspects of quantum discord see
also \cite{QD_R1, QD_R2, QD_R3, QD_Ficek, QD_Zeno,QD_2N,QD_X_Adesso}.

In the present paper we analyze a large class of two-qudit states
introduced in \cite{Circulant1} called circulant states (see also
\cite{Circulant2}). Construction of these states is based on a
certain decomposition of the total $d^2$-dimensional Hilbert space
into $d$ mutually orthogonal $d$-dimensional subspaces. A density
matrix $\rho$ representing a circulant state is a convex combination
of density matrices supported  on different subspace. Interestingly,
circulant states provide natural generalization of so called
$X$-states of two qubits. In the present paper we address a question
when a quantum discord for a circulant state does vanish.

The paper is organized as follows: in Section~\ref{DISC} we recall
basis definitions related to quantum discord and formulate necessary
and sufficient condition for the vanishing discord. Section
\ref{Circulant} introduces the construction of circulant states in
$\mathbb{C}^d \ot \mathbb{C}^d$ and analyzes the case of discord
zero. Section \ref{SYM} discusses several well known examples of
circulant states invariant under then action of the symmetry group
-- a subgroup of the unitary group $U(d)$.  In Section \ref{BELL} we
investigate special class of circulant states -- so called Bell
diagonal states \cite{Bell-2,Bell-5} -- which play
important role in the theory of quantum entanglement. To simplify
our discussion we consider only the case when $d$ is prime. Prime
dimensions already appeared for example in the discussion of
mutually unbiased basis \cite{MUB1}.
 It turns out that in this case the discussion
considerably simplifies. The general case may we analyzed in the
same way. However, the corresponding analysis is technically much
more involved. We illustrate our discussion with several examples
and conclude in the last Section.


\section{Quantum discord}  \label{DISC}


Consider a density operator $\rho$ of a composite quantum system in
Hilbert space $\mathcal{H}_{AB}=\mathcal{H}_A \otimes
\mathcal{H}_B$. The total amount of correlations in a bipartite
state $\rho$ is quantified by quantum mutual information:
\begin{equation}
\label{}
    \mathcal{I}(\rho) = S(\rho_A) + S(\rho_B) - S(\rho) \ ,
\end{equation}
where $\rho_{A}$ and $\rho_B$ are reduced density matrices in
$\mathcal{H}_A$ and $\mathcal{H}_B$ respectively and $S(\sigma)= -
{\rm tr}(\sigma \log\sigma)$ stands for the von Neumann entropy of
the density operator $\sigma$. Note, that mutual information may be rewritten as
follows
\begin{equation}\label{}
    \mathcal{I}(\rho) = S(\rho_B)  - S(\rho_{B|A}) \ ,
\end{equation}
where $\rho_{B|A}$ denotes a state of subsystem $B$ given
measurement on subsystem $A$ and $S(\rho_{B|A})$ is a quantum
conditional entropy. Let us introduce a local measurement on $A$
part defined by a collection of one-dimensional projectors
$\{\Pi_k\}$ in $\mathcal{H}_A$ satisfying $\Pi_1 + \Pi_2 + \ldots =
\mathbb{I}_A$. Different outcomes of this measurement are labeled by
`$k$'. The state of part $B$ after the measurement on part $A$, when
the outcome corresponding to $\Pi_k$ has been detected, is given by
\begin{equation}
    \rho_{B|k} = tr_A [\frac{1}{p_k} (\Pi_k \otimes \mathbb{I}_B)\rho (\Pi_k \otimes
    \mathbb{I}_B)\ ],
\end{equation}
where $p_k = {\rm tr}[\rho_{B|k} (\Pi_k\otimes \mathbb{I}_B)]$. The
entropies $S(\rho_{B|k})$ weighted by probabilities $p_k$ yield to
the conditional entropy of part $B$ given the complete measurement
$\{\Pi_k\}$ on the part $A$
\begin{equation}
    S(\rho_B|\{\Pi_k\}) = \sum_k p_k S(\rho_{B|k})\, .
\end{equation}
This means that the corresponding measurement-introduced mutual
information is
\begin{equation}\label{}
    \mathcal{I}(\rho_B|\{\Pi_k\}) = S(\rho_B) - S(\rho_B|\{\Pi_k\}) \, .
\end{equation}
By optimizing over all possible measurements $\{\Pi_k\}$ on part $A$
one has
\begin{equation}\label{}
    \mathcal{C}_{A}(\rho) = \sup_{\{\Pi_k\}} \mathcal{I}(\rho_B|\{\Pi_k\})\, .
\end{equation}
This quantity has been given an interpretation as measure of
classical correlations.

Although this two quantities, $\mathcal{I}(\rho)$ and
$\mathcal{C}_A(\rho)$, are equivalent for classical systems, in
quantum domain they, in general, do not coincide. The difference
\begin{equation}
\label{D_A}
    \mathcal{D}_{A}(\rho) =  \mathcal{I}(\rho) - \mathcal{C}_A(\rho)
\end{equation}
defines a new quantity, quantum discord, which is a measure of
quantum correlations in a quantum state $\rho$.

Evidently, the above definition is not symmetric with respect to
parties $A$ and $B$. One can swap the role of $A$ and $B$,
introducing a collection of one-dimensional $\Pi^B_\alpha$
projectors in $\mathcal{H}_B$ satisfying $\Pi^B_1 + \Pi^B_2 + \ldots
= \mathbb{I}_B$. Then one gets an analogous definition for discord
of a composite system when part $B$ is measured
\begin{equation}
\label{D_B}
    \mathcal{D}_{B}(\rho) =  \mathcal{I}(\rho) - \mathcal{C}_B(\rho),
\end{equation}
where
\begin{equation}
    \mathcal{C}_{B}(\rho) = \sup_{\{\Pi^B_\alpha\}} \mathcal{I}(\rho|\{\Pi^B_\alpha\})\, .
\end{equation}

Quantum discord $\mathcal{D}_A(\rho)$
and $ \mathcal{D}_B(\rho)$, is always non-negative. Although for all
states with the same reduced density matrices $\mathcal{D}_A(\rho) =
\mathcal{D}_B(\rho)$, this in general is not the case. Moreover, on pure
states, quantum discord coincides with the von Neumann entropy of
entanglement $S(\rho_A) = S(\rho_B)$.
One shows that $\mathcal{D}_A(\rho)=0$ (so called classical-quantum states)
if and only if there exists an orthonormal basis $|k\rangle $ in
$\mathcal{H}_A$ such that
\begin{equation}
    \rho = \sum_k p_k\, |k\rangle \langle k| \otimes \rho^{(B)}_k \, ,
\end{equation}
where $\rho^{(B)}_k$ are density matrices in $\mathcal{H}_B$.
Similarly, $\mathcal{D}_B(\rho)=0$ (quantum-classical states), if and only if there exists an
orthonormal basis $|\alpha \rangle $ in $\mathcal{H}_B$ such that
\begin{equation}\label{}
    \rho = \sum_\alpha q_\alpha\, \rho^{(A)}_\alpha \otimes |\alpha\rangle \langle \alpha| \, ,
\end{equation}
where $\rho^{(A)}_\alpha$ are density matrices in $\mathcal{H}_A$. It is clear that if
$\mathcal{D}_A(\rho)=\mathcal{D}_B(\rho)=0$, then $\rho$ is diagonal
in the product basis $|k\rangle  \otimes |\alpha \rangle $ and hence
\begin{equation}\label{}
    \rho = \sum_{k,\alpha} \lambda_{k\alpha}\, |k\rangle \langle k| \otimes |\alpha\rangle \langle \alpha| \, ,
\end{equation}
is entirely represented by the classical joint probability
distribution $\lambda_{k\alpha}$. Such states are called completely
classical.

States with a positive quantum discord do contain non-classical
correlations  even if they are separable. Hence nonvanishing quantum
discord indicates a kind of  quantumness encoded in a separable
mixed state.


Consider now states with vanishing quantum discord \cite{QD_Zero+Geo, ZeroQD_Modi,ZeroQD_Datta}.
Take two arbitrary orthonormal basis $|\eA_i\>$ and $|\eB_\alpha\>$ in
$\mathcal{H}_A$ and $\mathcal{H}_B$,  respectively. An arbitrary state
$\rho_{AB}$ in $\mathcal{H}_A \ot \mathcal{H}_B$ may be written as follows
\begin{equation}\label{}
    \rho_{AB} = \sum_{i,j} \eA_{ij} \ot \rB_{ij}\ ,
\end{equation}
or
\begin{equation}\label{}
    \rho_{AB} = \sum_{\alpha,\beta} \rA_{\alpha\beta}\ot \eB_{\alpha\beta}\  ,
\end{equation}
where
$$\eA_{ij} = |\eA_i\>\<\eA_j|\ ,\ \ \    \eB_{\alpha\beta} = |\eB_\alpha\>\<\eB_\beta|\ , $$
defines orthonormal basis in $\mathcal{B}(\mathcal{H}_A)$ and $\mathcal{B}(\mathcal{H}_B)$, respectively, and
$$\rB_{ij} \in \mathcal{B}(\mathcal{H}_B)\ , \ \ \
\rA_{\alpha\beta} \in \mathcal{B}(\mathcal{H}_A)\ . $$

\begin{tw} $D_A(\rho_{AB})=0$ iff $\rA_{\alpha\beta}$ are simultaneously diagonalizable.
Similarly, $D_B(\rho_{AB})=0$ iff $\rB_{ij}$ are simultaneously diagonalizable.
\end{tw}
Now it is well known, that  if $\rB_{ij}$ are simultaneously diagonalizable then they  mutually  commute, i.e.
\begin{equation}\label{}
    [ \rB_{ij},\rB_{kl}]=0\ .
\end{equation}
Note, that
\begin{equation}\label{}
    \rB_{ij} = \rho^{(B)\dagger}_{ji}\ ,
\end{equation}
and hence $ [ \rB_{ij},\rB_{ji}]=0$ implies that all $\rB_{ij}$ are normal (clearly,
the diagonal blocks are $\rB_{ii}$ are Hermitian and hence normal as well).

\begin{cor}  \label{COR_1}
If at least one off-diagonal  block $\rB_{ij}$ ($\rA_{\alpha\beta}$) is not normal then $D_B(\rho_{AB}) >0$ ($D_A(\rho_{AB})>0$).
\end{cor}

Normal matrices are simultaneously diagonalizable  if and only if they  mutually  commute
and hence

\begin{cor} \label{COR_2}
$D_A(\rho_{AB})=0$ iff $\rA_{\alpha\beta}$ mutually commute.
Similarly, $D_B(\rho_{AB})=0$ iff $\rB_{ij}$ mutually commute.

\end{cor}


\section{Circulant states in $\mathbb{C}^d \ot \mathbb{C}^d$}
\label{Circulant}

Let $\{|e_0\rangle,\cdots,|e_{d-1}\rangle \}$
denotes an orthonormal basis in $\mathbb{C}^d $. One introduces
shift operator $S:\mathbb{C}^d \rightarrow \mathbb{C}^d$ defined as
follows
\begin{equation}
\label{Shift} S |e_n\rangle=|e_{n+1}\rangle, \;\; (\text{mod } d).
\end{equation}
Now, let us define
\begin{equation}
\Sigma_0 =\text{span} \{|e_0\rangle \otimes
|e_0\rangle,\cdots,|e_{d-1}\rangle \otimes |e_{d-1}\rangle  \}
\end{equation}
and
\begin{equation}
\label{ } \Sigma_n=(\mathbb{I} \otimes S^n)\Sigma_0,
\end{equation}
for $n=1,\cdots,d-1$. It is clear that $d$-dimensional subspaces $\Sigma_n$ and
$\Sigma_m$ are mutually orthogonal for $m \neq n$ and hence one has the following
direct sum decomposition
\begin{equation}
\label{ } \Sigma_0 \oplus  \cdots \oplus \Sigma_{d-1} =\mathbb{C}^d
\otimes \mathbb{C}^d.
\end{equation}

Now, consider a class of states living in $\mathbb{C}^d \otimes
\mathbb{C}^d$ that may be written as a direct sum
\begin{equation}
\rho= \rho_0 \oplus \ldots \oplus \rho_{d-1}\ ,
\end{equation}
where each $\rho_n$ are supported on  $\Sigma_n$, that is,
\begin{equation}\label{}
    \rho_n = \sum_{i,j=1}^{d-1} a_{ij}^{(n)}
e_{ij}\otimes S^n e_{ij} S^{\dagger n}\ ,
\end{equation}
where $[a_{ij}^{(n)}]$ is a $d\times d$ semi-positive matrix, for
$n=0,\cdots,d-1$. Normalization of $\rho$ implies following
condition for matrices $a^{(n)}$
\begin{equation}
\label{ } {\rm Tr}(a^{(0)}+\cdots+a^{(d-1)})=1.
\end{equation}
These states were called circulant \cite{Circulant1,Circulant2} due to the cyclic structure of
the shift operator $S$.  The above construction defines natural
generalization of the well known $X$-states. Indeed, for
$d=2$ one obtains
\begin{equation} \label{2x2}
\rho=\left(\begin{array}{cc|cc}
a_{00} &.  & . & a_{01}\\
 . & b_{00} & b_{01} & . \\
 \hline
 . & b_{10} & b_{11} & . \\
 a_{10} &.  & . & a_{11}\\
\end{array} \right)
\end{equation}
where to make the picture more transparent we replaced all zeros by
dots  and introduced two matrices $a:=a^{(0)}$, $b:=a^{(1)}$. For
$d=3$ the structure of a circulant state reads as follows
\begin{equation}   \label{3x3}
\rho=\left(\begin{array}{ccc|ccc|ccc}
a_{00} &.  & . & . & a_{01}&.&.&.&a_{02}\\
 . & b_{00} &.&.&.& b_{01} & b_{02}&.&. \\
 .&.&c_{00}&c_{01}&.&.&.&c_{02}&.\\
 \hline
 .&.&c_{10}&c_{11}&.&.&.&c_{12}&.\\
a_{10} &.  & . & . & a_{11}&.&.&.&a_{12}\\
 . & b_{10} &.&.&.& b_{11} & b_{12}&.&. \\
\hline
 . & b_{20} &.&.&.& b_{21} & b_{22}&.&. \\
 .&.&c_{20}&c_{21}&.&.&.&c_{22}&.\\
a_{20} &.  & . & . & a_{21}&.&.&.&a_{22}\\

\end{array} \right)
\end{equation}
where $a:=a^{(0)}$, $b:=a^{(1)}$, $c:=a^{(2)}$. Actually, it turns
out that many well known examples of quantum states of composite
systems belong to the class of circulant states: the most prominent
are Werner state, isotropic state, states invariant under the local
action of the unitary group $U(d)$ and many others
\cite{Circulant1}. In Section \ref{BELL} we analyze an interesting
subclass of circulant states -- Bell diagonal states.

Let $\Pi^{(AB)}_n$ be an orthogonal projector onto $\Sigma_n$, that
is
\begin{equation}\label{}
\Pi^{(AB)}_n = \sum_{i=0}^{d-1}\, e_{ii} \ot e_{i+n,i+n}\ .
\end{equation}
We add a superscript ``AB'' to emphasize that $\Pi^{(AB)}_n$ is a
non-local  (or rather non-separable)  projector. Note, that
$\rho_{AB}$ is circulant if and only if
\begin{equation}\label{}
\rho = \sum_{n=0}^{d-1} \Pi^{(AB)}_n  \rho  \Pi^{(AB)}_n\ .
\end{equation}
Now, let us look for a circulant states with vanishing quantum
discord. One easily finds for the corresponding blocks:
\begin{equation}\label{}
    \rB_{ij} = \sum_{n=0}^{d-1} a^{(n)}_{ij} e_{i+n,j+n}\ ,
\end{equation}
and
\begin{equation}\label{}
    \rA_{ij} = \sum_{n=0}^{d-1} a^{(n)}_{i-n,j-n} e_{i-n,j-n}\ .
\end{equation}
Due to the Corollary \ref{COR_1} the necessary condition for
$D_B(\rho)=0$ ($D_A(\rho)=0$) is that $\rB_{ij}$ ($\rA_{ij}$) are
normal. To simplify our analysis we shall consider only prime
dimension $d$.

\begin{prop} \label{PRO_1}
If $d$ is prime, then off-diagonal blocks $\rB_{ij}$ are normal iff
\begin{equation}\label{NB}
|a_{ij}^{(n)}|=|a_{ij}^{(0)}|\ ,
\end{equation}
for  $i, j, n=0,1,2$. The off-diagonal blocks $\rA_{ij}$ are normal
iff
\begin{equation}\label{NA}
|a_{ij}^{(n)}|=|a_{i+n,j+n}^{(0)}|\ ,
\end{equation}
for  $i, j, n=0,1,2$.
\end{prop}
If $d$ is not prime then (\ref{NB}) and (\ref{NA}) are sufficient
but not necessary to guarantee that $\rB_{ij}$ and $\rA_{ij}$ are
normal. Note, that if $d$ is not prime, then off-diagonal blocks may
display sub-block structure corresponding to the decomposition $d =
d_1^{p_1} \ldots d_r^{p_r}$, with $d_1,\ldots,d_r$ prime numbers. We
shall not analyze this situation in the present paper.

Note, that if all $a^{(n)}$  are diagonal, then $\rA_{ij}=\rB_{ij}= 0$ and
the corresponding circulant state is purely classical. Suppose now that at least
one matrix $a^{(n)}$  is not diagonal.

\begin{prop} \label{PRO_2}
If the off-diagonal blocks $\rB_{ij}$ are normal, then
 $   [ \rB_{kk},\rB_{ij}] = 0$,
iff
\begin{equation}\label{DB}
a_{kk}^{(n)} = a_{kk}^{(0)}\ .
\end{equation}
Similarly, If the off-diagonal blocks $\rA_{ij}$ are normal, then
 $   [ \rA_{kk},\rA_{ij}] = 0$,
iff
\begin{equation}\label{DA}
a_{kk}^{(n)} = a_{k+n,k+n}^{(0)}\ .
\end{equation}
\end{prop}
It should be stressed that conditions (\ref{NB}) and (\ref{DB}) are
necessary for $D_B(\rho)=0$. Similarly, conditions (\ref{NA}) and
(\ref{DA}) are necessary for $D_A(\rho)=0$.

Now, let us formulate sufficient conditions. Let $V$ be a unitary
operator
\begin{equation}
V=\sum_{n=0}^{d-1} e_{nn} e^{i \phi_n},
\end{equation}
with $\phi_0=0$ (the global phase would play no role in what
follows). The main result of this paper consists in the following

\begin{tw} \label{tw_DA}
Assume that at least
one matrix $a^{(n)}$  is not diagonal and $d$ is prime.
\begin{enumerate}
\item  $D_A(\rho)=0$, if and only if
\begin{equation}
a^{(k)}=(VS ^{\dagger})^k a^{(0)} (SV^{\dagger})^k,
\end{equation}
for $k=1,\cdots, d-1$.

\item $D_B(\rho)=0$, if and only if
\begin{equation}
\label{V_DB} a^{(k)}=S^{\dagger(k-1)}(VS)^{(k-1)} V
a^{(0)}V^{\dagger}(VS)^{\dagger(k-1)} S^{k-1},
\end{equation}
for $k=1,\cdots, d-1$.
\end{enumerate}
\end{tw}
Hence, discord zero circulant state is fully characterized by a single
matrix $a^{(0)} \geq 0$ and a unitary operator $V$.  In particular taking
$V=\mathbb{I}$ one obtains
\begin{equation}
a^{(k)}= S^{\dagger k} a^{(0)} S^{ k} ,
\end{equation}
for $D_A(\rho)=0$, and
\begin{equation}
a^{(k)}=  a^{(0)} ,
\end{equation}
for $D_B(\rho)=0$.


\begin{ex}
 \em{
In two-qubit case a circulant state is given by (\ref{2x2}). Interestingly,
if (\ref{2x2}) is not diagonal, then $X$-state with vanishing discord
is fully characterized by Propositions \ref{PRO_1} and \ref{PRO_2}:

\begin{enumerate}

\item $X$-state has vanishing $D_A$, iff
\begin{equation}
a_{00} =  b_{11}, \ \ \
a_{11} = b_{00} \ ,
\end{equation}
and
$|a_{01}| = |b_{01}|$.

\item $X$-state has vanishing $D_B$, iff
\begin{equation}
a_{00} =  b_{00}, \ \ \
a_{11} = b_{11} \ ,
\end{equation}
and
$|a_{01}| = |b_{01}|$.

\item $X$-state is purely classical, i.e. $D_A(\rho)=D_B(\rho)=0$, iff
\begin{equation}
a_{00} =  b_{00} = a_{11} = b_{11} = \frac 14 \ ,
\end{equation}
and
$|a_{01}| = |b_{01}|$.

\end{enumerate}
These results were already derived in \cite{QD_Luo,QD_X,QD_X2,ZeroQD_2N}.
}
\end{ex}

\begin{ex}
\em{ Consider now a circulant state in $3\ot 3$ defined in (\ref{3x3}).
Assume that a matrix $a_{ij}$ is not diagonal.

\begin{enumerate}

\item $D_A(\rho)=0$, iff the matrices $b_{ij}$ and $c_{ij}$ are defined by

\begin{displaymath}
b=\left(\begin{array}{ccc}
a_{11} & a_{12} e^{i \varphi_1} & a_{10} e^{i(\varphi_1+\varphi_2)} \\
a_{21} e^{-i \varphi_1} &  a_{22}  & a_{20} e^{i\varphi_2} \\
a_{01} e^{-i(\varphi_1+\varphi_2)} & a_{02} e^{-i\varphi_2} & a_{00}
\end{array}\right) \ ,
\end{displaymath}
and
\begin{displaymath}
c=\left(\begin{array}{ccc}
a_{22} & a_{20} e^{i(\varphi_1+\varphi_2)} & a_{21} e^{i\varphi_2} \\
a_{02} e^{-i(\varphi_1+\varphi_2)} &  a_{00}  & a_{01} e^{-i\varphi_1} \\
a_{12} e^{-i\varphi_2} & a_{10} e^{i\varphi_1} & a_{11}
\end{array}\right) .
\end{displaymath}

\item $D_B(\rho)=0$, iff the matrices $b_{ij}$ and $c_{ij}$ are defined by

\begin{displaymath}
b=\left(\begin{array}{ccc}
a_{00} & a_{01} e^{-i\varphi_1} & a_{02} e^{-i(\varphi_1+\varphi_1)} \\
a_{10} e^{i \varphi_0} &  a_{11}  & a_{12} e^{-i\varphi_2} \\
a_{20} e^{i(\varphi_1+\varphi_2)} & a_{21} e^{i\varphi_1} &  a_{22}
\end{array}\right) \ ,
\end{displaymath}
and
\begin{displaymath}
c=\left(\begin{array}{ccc}
a_{00} & a_{01} e^{-i(\varphi_1+\varphi_2)} & a_{02} e^{-i\varphi_2} \\
a_{10} e^{i(\varphi_1+\varphi_2)} &  a_{11}  & a_{12} e^{i\varphi_1} \\
a_{20} e^{i\varphi_2} & a_{21} e^{-i\varphi_1} &  a_{22}
\end{array}\right) \ .
\end{displaymath}

\item If $D_A(\rho)=D_B(\rho)$, then
\begin{equation}\label{}
    a^{(n)}_{ii} = \frac{1}{9} \ ,
\end{equation}
and
\begin{equation}\label{}
   |a^{(n)}_{ij}| = {\rm const.}
\end{equation}
for $i\neq j$.

\end{enumerate}

 }
\end{ex}
More generally one has the following

\begin{tw}
\label{tw_DA&DB} A two-qudit circulant state $\rho$ living in
$\mathbb{C}^d \otimes \mathbb{C}^d$, where $d$ is a prime number, is
completely classical, i.e. $D_A(\rho)=0$ and $D_B(\rho)=0$, if and
only if
\begin{equation}
a_{ii}^{(0)}=\frac{1}{d^2},
\end{equation}
for $i=0,\cdots,d-1$, and the off-diagonal elements
\begin{equation}
|a_{ij}^{(0)}|= {\rm const.}
\end{equation}
The remaining matrices $a^{(n)}$ are defined as follows
\begin{equation}
\label{V_DAB} a^{(k)}=(VS ^{\dagger})^k a^{(0)} (SV^{\dagger})^k.
\end{equation}
\end{tw}

\section{Circulant symmetric states} \label{SYM}

Interestingly several classes of symmetric states, like e.g. Werner or isotropic states,
belong to the class of circulant states.
Let $G $ be a subgroup of the unitary group $U(d)$. A bipartite operator  $A$ living
in $\mathbb{C}^d \ot \mathbb{C}^d$ is $G \ot G$--invariant \cite{Voll,PRA-I-II} if
\begin{equation}\label{}
    U \ot U \rho = \rho U \ot U\ ,
\end{equation}
for all $U \in G$. Note, that if $A$ is $G \ot G$--invariant
then its partial transposition $A^\Gamma$ is $U \ot \overline{U}$--invariant, where
$\overline{U}$ denotes complex conjugation with respect to a fixed basis in $\mathbb{C}^d$.

If $G=U(d)$ then  $G \ot G$--invariant state -- Werner state -- is given by
\begin{equation}
\rho_W=\frac{1-\lambda}{d^2}\, \mathbb{I}\ot \mathbb{I} + \frac{\lambda}{d}\, \mathbb{F},
\end{equation}
where $\mathbb{F}$ is the flip operator defined by $\mathbb{F}=\sum_{i,j=0}^{d-1} e_{ij}\otimes e_{ji}$.
It is a circulant state and the corresponding matrices $a^{(n)}$ read as follows
\[a^{(0)}=
\begin{cases}
0 & , i \neq j,\\
\frac{\lambda}{d}+ \frac{1-\lambda}{d^2} & , i=j
\end{cases} \]

\[a^{(k)}=
\begin{cases}
\frac{\lambda}{d} & , j=i+k,\\
\frac{1-\lambda}{d^2}& , i=j
\end{cases} \]
Hence $D_A(\rho_W)=D_B(\rho_W)=0$ only if $\lambda=0$.

Similarly an isotropic states which is invariant under $G \ot \overline{G}$ is defined by
\begin{equation}
\rho_I=\frac{1-\lambda}{d^2}\, \mathbb{I}\ot \mathbb{I} + \lambda\, P_{d}^{+},
\end{equation}
where
\begin{equation}\label{}
P^+_d = \frac 1d \sum_{i,j=0}^{d-1} e_{ij}\otimes e_{ij}\ .
\end{equation}
One finds
\[a^{(0)}=
\begin{cases}
\frac{\lambda}{d} & , i \neq j,\\
\frac{\lambda}{d}+ \frac{1-\lambda}{d^2}& , i=j
\end{cases} \]

\[a^{(k)}=
\begin{cases}
0 & , i \neq j,\\
\frac{1-\lambda}{d^2}& , i=j,
\end{cases} \]
for $k=1,\cdots, d-1.$ Again $D_A(\rho_I)=D_B(\rho_I)=0$ only if $\lambda=0$.
This results agree with \cite{QD_GeoL} where geometric discord for two-qudit Werner and Isotropic states was calculated.

Consider now $G$ consisting of real unitary operators from $U(d)$ (again in fixed basis in $\mathbb{C}^d$).
It turns out that $G=O(d)$ \cite{Voll,PRA-I-II}. One shows that $O(d)\ot O(d)$--invariant state
has the following form
\begin{equation}\label{}
    \rho = a \widetilde{\mathcal{P}}_0 + b   \widetilde{\mathcal{P}}_1 + c\widetilde{\mathcal{P}}_2  ,
\end{equation}
with $a+b+c=1$ and $a,b,c\geq0$. Normalized projectors $\widetilde{\mathcal{P}}_0$ are defined as follows
$ \widetilde{\mathcal{P}}_k =  {\mathcal{P}}_k / {\rm Tr} {\mathcal{P}}_k$, where
\begin{equation}\label{}
{\mathcal{P}}_0 = Q^+ - P^+\ , \ \  {\mathcal{P}}_1 = Q^- \ , \ \  {\mathcal{P}}_2 = P^+_d\ ,
\end{equation}
and $Q^+$, $Q^-$ are projectors onto the symmetric and antisymmetric subspaces in $\mathbb{C}^d \ot \mathbb{C}^d$, that is,
\begin{equation}\label{}
    Q^\pm = \frac 12 ( \mathbb{I}\ot \mathbb{I} \pm \mathbb{F}) \ .
\end{equation}
Again, one easily shows that for $d > 2$ a symmetric state $\rho$ has vanishing discord if and only if $\rho$ is maximally mixed.
Interestingly for $d=2$ a class of symmetric discord zero states is nontrivial.
The density matrix has the following form
\begin{equation} \label{}
\rho= \frac 14 \left(\begin{array}{cc|cc}
a + 2c &.  & . & 2c-a\\
 . & a+2b & a-2b & . \\
 \hline
 . & a-2b & a+2b & . \\
 2a-c &.  & . & a+2c\\
\end{array} \right) \ ,
\end{equation}
and hence it belongs to the class of $X$-states. It is well known that $\rho$ is separable iff
$b,c \leq 1/2$. Note that $\rho$ has vanishing discord iff $b=c$ (see Fig. 1). Note, that the simplex in the $bc$--plane is defined by three vertices $\widetilde{\cal P}_k$: separable $\widetilde{\mathcal{P}}_0$ and entangled $\widetilde{\mathcal{P}}_1$ and $\widetilde{\mathcal{P}}_2$. Interestingly, $\widetilde{\mathcal{P}}_0$ is not only separable but even purely classical.

\begin{figure}[h]
\begin{center} \includegraphics[width=7.5cm]{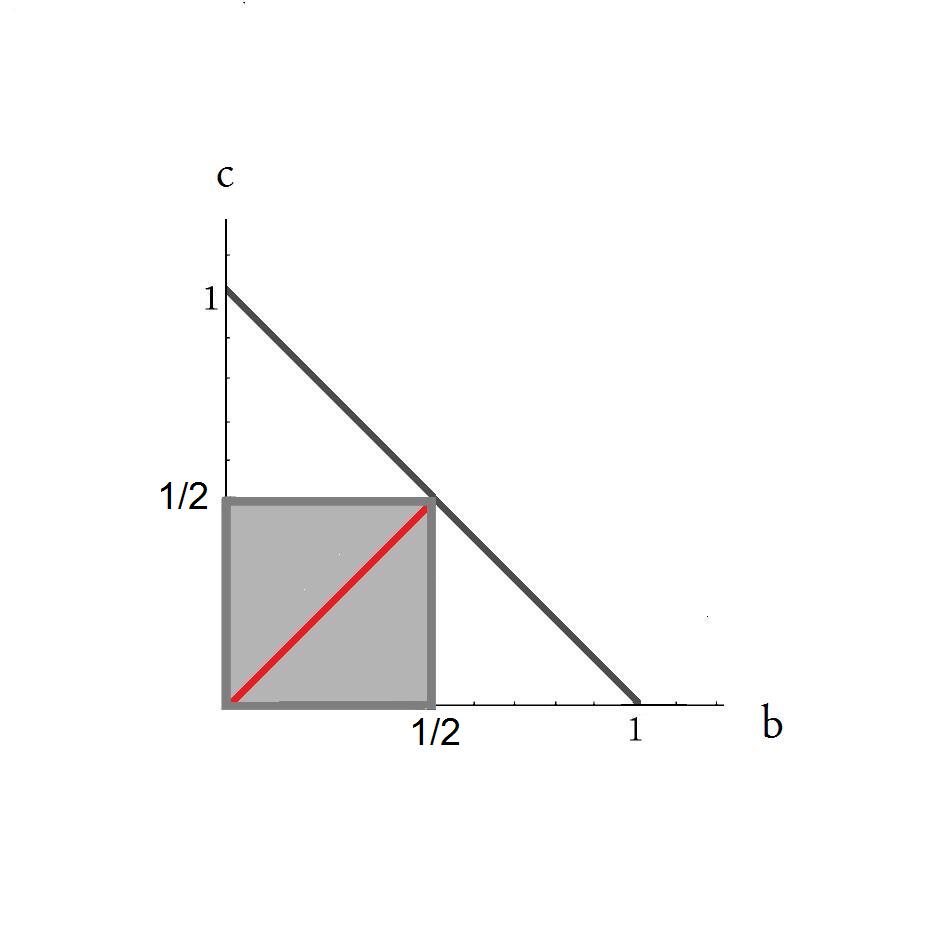}
\end{center}
 \caption{A simplex of states with orthogonal symmetry for d=2. Separable states form a gray square while zero-discord states are represented by the red line.} \label{rys1}
\end{figure}

Finally, let $G$ be a maximally commuting subgroup of $U(d)$ (again with respect to a fixed basis in $\mathbb{C}^d$).
It was shown \cite{PPT-nasza} that $G \ot G$-invariant state has the following form
\begin{equation}\label{}
    \rho = \sum_{i,j=0}^{d-1} a_{ij} e_{ij} \ot e_{ij} + \sum_{i\neq j} d_{ij} e_{ii} \ot e_{jj}\ ,
\end{equation}
where $a_{ij}$ is a $d\times d$ positive matrix and $d_{ij}$ are non-negative numbers.
Evidently, $\rho$ defines a circulant state with $a^{(0)}=a$ and $a^{(k)}$ are diagonal for $k>0$.
Interestingly,
this class of symmetric states is characterized by a simple PPT condition, namely, $\rho$ is PPT iff
\begin{equation}\label{}
    |a_{ij}|^2 \leq d_{ij} d_{ji} \ , \ \ \ i\neq j\ .
\end{equation}
However, the general condition for separability is not known \cite{PPT-nasza}. Note that $\rho$ has vanishing discord
iff $a_{ij}=0$ for $i \neq j$, that is, $\rho$ is diagonal.


\section{Generalized Bell diagonal states} \label{BELL}

In this section we analyze an important subclass of circulant
states. Let us introduce $d^2$ maximally entangled projectors
defined by
\begin{equation}
P_{mn}=(\mathbb{I} \otimes U_{mn}) P^+_d (\mathbb{I} \otimes
U_{mn}^\dagger),
\end{equation}
where $U_{mn}$ are unitary matrices defined as follows:
\begin{equation}  \label{basic}
U_{mn} |e_k\rangle = \lambda ^{mk} S^{n} |e_{k}\rangle,
\end{equation}
with $S$ being the shift operator defined in (\ref{Shift}), $P^+_d$
- projector on maximally entangled state, and
$\lambda = e^{2 \pi i/d}$.

\begin{Rm} {\em
Actually, one may define a more general class of states based on a
class of `shift and multiply basis' of unitary matrices in
$\mathbb{C}^d$ defined as follows \cite{HAD}
\begin{equation}\label{had}
    U_{ij} |e_k\> = H^j_{ik} |e_{L(j,k)}\> \ ,
\end{equation}
where a set of complex numbers $H^j_{ik}$, and $L :  I_d \times I_d
\rightarrow I_d$ with $I_d := \{0,1,\ldots,d-1\}$, satisfy the
following conditions:

i) each $H^j$ is a Hadamard matrix,

ii) $L$ is a Latin square, i.e., the maps $k \rightarrow L(k; \ell)$
and $\ell \rightarrow L(k; \ell)$ are injective for every $\ell$. It
is clear that (\ref{basic}) defines a special example of
(\ref{had}). }
\end{Rm}

Consider simplex of Bell diagonal states defined by
\begin{equation}
\label{Bell_def} \rho=\sum_{m,n=0}^{d-1} p_{mn} P_{mn},
\end{equation}
where $p_{mn}\geq 0$ and $\sum_{m,n=0}^{d-1} p_{mn}=1$. It is
evident from the construction that Bell diagonals states belong to
the class of circulant states.  One can easily check that
corresponding  matrices $a^{(n)}$ have the following form
\begin{equation}
a^{(n)}_{ij}=\frac{1}{d}\sum_{m=0}^{d-1} p_{mn} \lambda^{m(i-j)},
\end{equation}
hence defining Bell diagonal state is equivalent with determining
$d^2$ coefficients $p_{mn}$. Let us notice that marginal density
matrices of $\rho$ are equal, $\rho_A=\rho_B=\mathbb{I}/d$ which
means that $D_A(\rho)=0$ if and only if $D_B(\rho)=0$. Hence,
whenever Bell diagonal states is classical with respect to one party
it is already completely classical.

Consider now $\pi_k \geq 0$ $(k=0,\ldots,d-1$) such that
\begin{equation}\label{}
    \sum_{k=0}^{d-1} \pi_k= \frac 1d\ .
\end{equation}
Using results of the previous
section one proves

\begin{tw}
\label{tw_Bell} A Bell diagonal state (\ref{Bell_def}), living in $\mathbb{C}^d
\otimes \mathbb{C}^d$, where $d$ is a prime number, is a
zero-discord state if and only if
\begin{equation}
p_{ik}=\pi_{i+k\alpha}  \;\; (\text{mod } d),
\end{equation}
for some $\alpha\in \{0,1,\ldots,d-1\}$.
\end{tw}

Hence, any Bell diagonal state is uniquely determined by a vector $\pi_k$ and
the number `$\alpha$'.
\begin{ex}
\em{ Density operator for two-qubit case is defined by the following
matrices $a^{(n)}$
\begin{equation}
a^{(n)}=\left(\begin{array}{cc}
x_n &y_n\\
 y_n & x_n \\
\end{array} \right),
\end{equation}
for $n=0,1$, where
\begin{equation}
x_n=\frac{1}{2}(p_{0n}+p_{1n}), \;\;\;
y_n=\frac{1}{2}(p_{0n}-p_{1n}),
\end{equation}
This state is classical if and only if $x_n=1/4$ and $y_1= \pm y_0$.
In terms of the probability matrix
\begin{equation*}\label{}
    p_{ij} = \left(\begin{array}{cc}
p_{00} & p_{01}\\
p_{10} & p_{11} \\
\end{array} \right),
\end{equation*}
one has
\begin{equation}\label{}
    \left(\begin{array}{cc}
\pi_0 & \pi_0\\
 \pi_1 & \pi_1 \\
\end{array} \right), \ \ \   \left(\begin{array}{cc}
\pi_0 & \pi_1\\
 \pi_1 & \pi_0 \\
\end{array} \right),
\end{equation}
corresponding to $\alpha=0$ and $\alpha=1$, respectively.

}

\end{ex}

\begin{ex}
\em{ A two-qutrit Bell diagonal state is defined by matrices
\begin{equation}
a^{(n)}=\left(\begin{array}{ccc}
x_n &z_n&\overline{z_n}\\
\overline{z_n}&x_n & z_n \\
z_n&\overline{z_n}&x_n
\end{array} \right),
\end{equation}
for $n=0,1,2$, where
\begin{equation}
x_n=\frac{1}{3}(p_{0n}+p_{1n}+p_{2n})
\end{equation}
and
\begin{equation}
z_n=\frac{1}{3}(p_{0n}+\overline{\lambda} p_{1n}+ \lambda p_{2n}).
\end{equation}
This state is classical if and only if diagonal elements  $
x_n=1/9$ and  the off-diagonal elements fulfill one of the
following conditions
\begin{equation}
z_n=\lambda^{n \alpha}z_0\ ,\ \ \ n=1,2\ ,
\end{equation}
where $\alpha \in \{0,1,2\}$.
In terms of the probability matrix $p_{ij}$ one has
$$
\left(\begin{array}{ccc}
\pi_0 & \pi_0 & \pi_0\\
\pi_1 & \pi_1 & \pi_1\\
\pi_2 & \pi_2 & \pi_2\\
\end{array} \right), \left(\begin{array}{ccc}
\pi_0 & \pi_1 & \pi_2\\
\pi_1 & \pi_2 & \pi_0\\
\pi_2 & \pi_0 & \pi_1\\
\end{array} \right), \left(\begin{array}{ccc}
\pi_0 & \pi_2 & \pi_1\\
\pi_1 & \pi_0 & \pi_2\\
\pi_2 & \pi_1 & \pi_0\\
\end{array} \right),    $$
for $\alpha=0$, $\alpha=1$ and $\alpha=2$, respectively.
}
\end{ex}


\section{Conclusions}

We analyzed a large class of two-qudit circulant states
which provide natural generalization of the celebrated $X$--states.
For prime dimension $d$ we formulated  necessary and sufficient conditions for vanishing discord.
It turns out that such states are fully characterized by a density operator living on one particular
subspace from the direct sum decomposition $\Sigma_0 \oplus \ldots \oplus \Sigma_{d-1}$.
Interestingly the class of circulant states contains several well known classes of symmetric states, i.e. states invariant under the local action of $U(d)$ or its subgroups. Finally, we characterized Bell diagonal states (another important class of circulant states) with vanishing quantum discord. This analysis generalizes the well known characterization of Bell diagonal states of two qubits.


\end{document}